# Role of Strain on the Coherent Properties of GaAs Excitons and Biexcitons

Brian L. Wilmer,[1] Daniel Webber,[2] Joseph M. Ashley,[1,†] Kimberley C. Hall,[2] and Alan D. Bristow[1,*]

[1]Department of Physics and Astronomy, West Virginia University, Morgantown WV 26506-6315, U.S.A.
[2]Department of Physics and Atmospheric Science, Dalhousie University, Halifax NS, Canada, B3H4R2

**Abstract**: Polarization-dependent two-dimensional Fourier-transform spectroscopy (2DFTS) is performed on excitons in strained bulk GaAs layers, probing the coherent response for differing amounts of strain. Biaxial tensile strain lifts the degeneracy of heavy-hole (HH) and light-hole (LH) valence states, leading to an observed splitting of the associated excitons at low temperature. Increasing the strain increases the magnitude of the HH/LH exciton peak splitting, induces an asymmetry in the off-diagonal interaction coherences, increases the difference in the HH and LH exciton homogenous linewidths, and increases the inhomogeneous broadening of both exciton species. All results arise from strain-induced variations in the local electronic environment, which is not uniform along the growth direction of the thin layers. For cross-linear polarized excitation, wherein excitonic signals give way to biexcitonic signals, the high-strain sample shows evidence of bound LH, HH, and mixed biexcitons.

## I. INTRODUCTION

Strain engineering provides a powerful method to tailor the electronic and optical properties of semiconductors, a feature that has been utilized in a variety of applications in semiconductor opto-electronics. For instance, the controlled use of strain has led to low-threshold lasers using strained quantum wells or quantum dot systems,[1–4] and fast field-effect transistors based on the SiGe/Si system.[5] Manipulation of strain may also lead to the development of polarization-entangled photon pairs through control of the relative size of the exciton and biexciton binding energies in quantum dot systems,[6] and provides a means to dynamically manipulate the spin-orbit interaction for possible spin-sensitive electronic devices.[7,8] For such applications, an accurate characterization of the influence of strain on the local and global optoelectronic properties in the semiconductor heterostructure is essential. Existing studies of strain in semiconductor systems have largely focused on its influence on the energetic locations of the heavy-hole (HH) and light-hole (LH) exciton resonances, detected using optical techniques such as photoluminescence, reflectivity and Raman scattering.[9–15] The exciton peak positions may be understood taking into account the hydrostatic and uniaxial stress components, the latter of which lifts the degeneracy in the valence bands. For quantum applications the entire response, including coherent properties must be understood.

Coherent nonlinear optical techniques, such as two-dimensional Fourier-transform spectroscopy (2DFTS), provide a more comprehensive picture of the excitonic response compared to linear optical techniques. 2DFTS may be exploited to gain further insight into the influence of strain on the coherent optical response of semiconductors. For instance, the ability to independently measure the levels of homogeneous and inhomogeneous broadening[16–21] may yield insight into varying local electronic environments tied to partial strain relaxation in thin film systems. 2DFTS has also provided a wealth of information about many-body interactions (MBI) in quantum well systems through the ability to isolate a range of quantum pathways from the excitons, their coupling to one another, and biexcitons.[22–32] The study of such effects in a bulk semiconductor system, in which the splitting between the exciton resonances is caused exclusively by strain, would provide insight into the influence of strain on MBI and the coherent response of the excitons.[33]

In this paper, 2DFTS is used to study bulk excitons in two GaAs layers with different thicknesses, such that the two samples exhibit different amounts of strain, resulting from being attached to sapphire disks and cooled. Comparison of rephasing 2DFTS spectra with collinear and cross-linear polarization configurations allows for the observation of the line shapes, center positions, homogeneous and inhomogeneous linewidths, and the coupling of LH, HH excitons and biexcitons. In addition to increasing the separation between the HH and LH excitons, these experiments show that increasing the amount of strain in the bulk semiconductor modifies the relative strengths of the off-diagonal exciton features, altering the spectrum from a coherent response typically seen in atomic vapors to one seen in quantum wells. A larger degree of inhomogeneous broadening is also observed in the sample with larger strain, attributed to variations in the local electronic environment that are likely tied to strain gradients along the growth direction of the heterostructure. Significant bound biexciton signatures are visible for cross-linear polarized excitation, due to the presence of few-body interactions. Overall, the results suggest that 2DFTS and the coherent response of GaAs excitons is highly sensitive to strain.

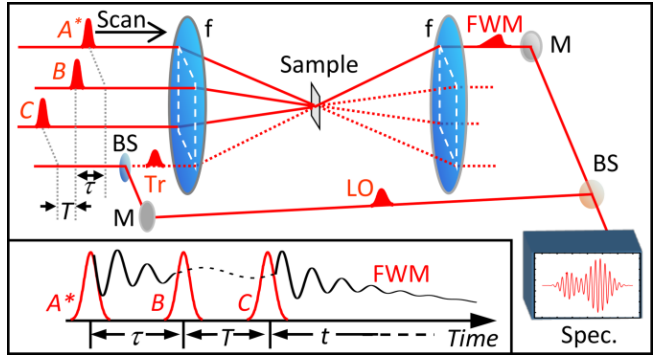

**Fig. 1** Experimental setup for the two-dimensional Fourier-transform spectroscopy with the sample at the focus. (f= lens, M = mirror, BS = beam splitter, FWM = four-wave-mixing signal, $\tau$ = period between pulses $A^*$ and $B$, $T$ = period between pulses $B$ and $C$, and t = period after pulse $C$ triggers the emission, Tr = tracer beam, traces the phase-matched direction, LO = local oscillator for spectral interferometry, and Spec is the imaging spectrometer with a charged coupled device). The inset shows the rephrasing excitation pulse sequence.

## II. EXPERIMENTAL

Samples of intrinsic GaAs are grown by molecular-beam epitaxy to have thicknesses of 300 nm and 800 nm along with etch stop layers. The samples are glued to sapphire disks and the GaAs substrates are removed for optical transmission measurements. All measurements were performed below 10 K in a cold finger cryostat. Cooling leads to uniform contraction of the sapphire. The different coefficients of thermal expansion for GaAs and sapphire results in two-dimensional in-plane tensile strain of the thin GaAs layers.[34] This in-plane strain necessarily causes slight compressive strain along the growth direction, in an attempt to maintain the GaAs unit cell volume. Strain is able to relax along the growth direction away from the point of attachment to the sapphire. The two samples possess different effective strain values, as well as differing variations in the local electronic environment due to differing amounts of strain relaxation along the growth direction.

Absorption and 2DFTS measurements were performed with resonant excitation. The 2DFTS setup, shown in Fig. 1, is described fully in previous work.[35,36] In brief, a mode-locked Ti:sapphire laser oscillator generates ~100 fs pulses at a repetition rate of 76 MHz. The pulses are split into four equal strength replicas, attenuated to an average power of 0.2 mW per beam and arranged on four corners of a box. Pulses $A^*$, $B$, and $C$ excite the sample and can be independently adjusted to control the delay times $\tau$ and $T$ (see inset of Fig. 1 for time sequence), while maintaining a phase stabilization of $< \lambda/100$. The three excitation pulses impinge the sample at non-normal incidence allowing for spatial selection of the emitted transient TFWM signal emitted in the background-free $k_{Sig} = -k_A + k_B + k_C$ phase-matching direction. This phase-matching condition selects the rephasing spectrum with the pulse ordering $A^*$, $B$, $C$, where the star denotes the phase-conjugated pulse.

The emitted signal is collected in a spectrometer with a cooled CCD camera following detection of the local oscillator (LO) pulse, which is routed around the sample to avoid unwanted pre-excitation or reference distortions. A time delay of several picoseconds between the LO and signal produces spectral interference fringes that allow the complex spectrum to be extracted. Spectral interferograms are collected as a function of delay time τ, which is stepped in precise phase-controlled increments to generate a series of measurements of delay time versus the emission photon energy.

A two-dimensional spectrum is acquired by performing a numerical Fourier transform with respect to the delay time τ. For rephasing spectra, the transform results in the absorption photon energy being negative and the diagonal of the spectrum is chosen to point down and to the right. For all spectra shown in this work, $T = 100$ fs to enforce strict time ordering and minimize contributions from incoherent energy relaxation, which can occur at large values of $T$.[37] Additionally, all excitation and emission pulses are set with either collinear (XXXX) or cross-linear (XYYX) polarization configurations, where the first three terms indicate the excitation and the fourth term indicates the emission polarization states.[24] Excitation irradiances are limited to ensure the response is in the $\chi^{(3)}$ nonlinear optical regime and the laser pulses are centered over the midpoint between the HH and LH excitons observed in the linear spectra.

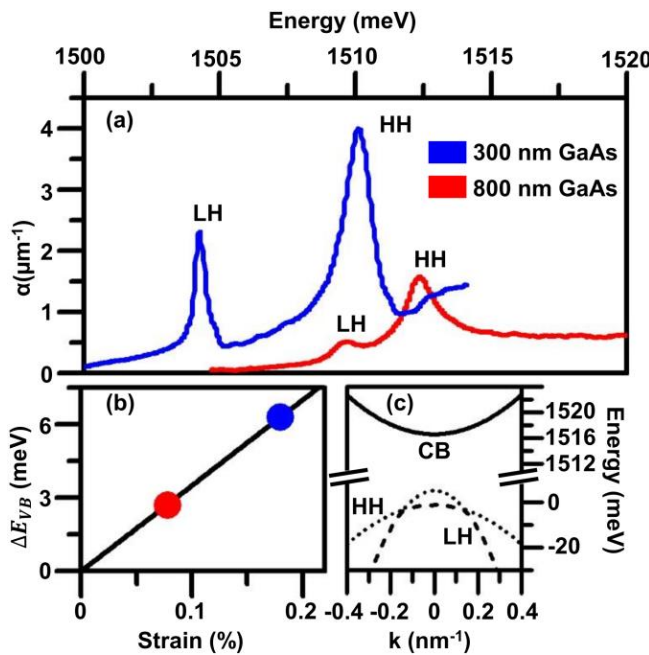

**Fig. 2** (a) Linear optical absorption spectra for a 300-nm (high strain) and 800-nm (low strain) GaAs layers revealing strain-split heavy- (HH) and light-hole (LH) excitons. (b) Effective strain (Δe) versus the exciton splitting (ΔE) for both samples. (c) Single-particle dispersions of the conduction, HH and LH valence bands.

### III. RESULTS AND DISCUSSION

Figure 2(a) shows the linear absorption of the 300-nm (blue) and 800-nm (red) GaAs layers. The spectra show sharp resonances due to the strain-split excitons and electron-hole continuum at higher energies. The thinner and thicker sample has a splitting between the exciton resonances of ~6 and ~3 meV. It is known that strain, among other variables, alters the band structure. Hydrostatic and biaxial tensile strain lift the degeneracy and puts the LH band energetically higher than the HH band. Perturbative strain linearly shifts the bands by an amount given by $E_{shift} = \Xi(\Delta a/a)$, where $\Xi$ is the deformation potential, $a$ is the lattice constant, and $\Delta a$ is the change in lattice constant due to stress.[38] Uniform biaxial in-plane strain lifts the degeneracy of the HH and LH valence bands, given by $\Delta E = 2|b|\Delta e$,[13,14] where $b$ is the shear deformation potential, and $\Delta e = e_{xx} - e_{zz}$ is the effective strain difference between the in-plane and out-of-plane directions. Using the literature value $b$ = -1750 meV,[39] effective strain differences of $0.18 \pm 0.01\%$ and $0.08 \pm 0.01\%$ are determined for the thinner and thicker GaAs layers respectively; see Fig. 2(b). In the figure, the error bars are smaller than the size of the data markers.

Conduction ($E_{cb}$), HH ($E_H$) and LH ($E_L$) bands are determined to have quadratic dispersions,[40] which are modified by linear shift terms at the Γ-point and band mixing away from $k = 0$. Final equations within an effective 1D model are simplified by assuming $e_{xx} = e_{yy}$, giving:

$$E_{cb} = \frac{\hbar^2}{2\,m_c}k^2 + E_g + a_c(2e_{xx} + e_{zz}), \tag{1}$$

$$E_L = -\frac{\hbar^2}{2\,m_e}\gamma_1 k^2 + a_v(2e_{xx} + e_{zz}) + \left|\frac{\hbar^2}{2\,m_L}\gamma_2 k^2 - b(e_{xx} - e_{zz})\right|, \tag{2}$$

and

$$E_H = -\frac{\hbar^2}{2\,m_e}\gamma_1 k^2 + a_v(2e_{xx} + e_{zz}) - \left|\frac{\hbar^2}{2\,m_H}\gamma_2 k^2 - b(e_{xx} - e_{zz})\right|. \tag{3}$$

Literature values for the effective masses of the conduction, LH, and HH bands are $m_C = 0.066 m_e$, $m_L = 0.094 m_e$ and $m_H = 0.34 m_e$ respectively,[41] where $m_e$ is the elementary electron mass. The unstrained low-temperature band gap is $E_g = 1518.9$ meV,[42] valence band deformation potential is $a_v = 1600\ meV$, conduction band deformation potential is $a_c =$ -7000 meV and the Luttinger parameters are $\gamma_1 = 6.98$, and $\gamma_2 = 2.06$.[3] The band gap is determined from the onset of the electron-LH continuum in Fig. 2(a) to be ~1512.8 meV. These values are used in conjunction with the observed exciton splitting, fitting parameters $e_{xx}$ and $e_{zz}$ to yield strains of 0.07% and -0.11% in the xy plane and z directions respectively for the high-strain sample. The resulting band edges are 1516.6 meV, 4.7 meV, and -1.5 meV for the conduction, LH and HH bands respectively. Equations (1)-(3) are plotted in Fig. 2(c) taking into account the literature and observed parameters. Away from the band Γ point, anti-crossings appear where the HH and LH bands are in close proximity. Although high wave vector regions of the dispersions are out of the accessible range of wave vectors used in the experiment. Within the light cone, near the Γ point, $E_L > E_H$, such that the lower energy spectral feature is the LH exciton. Similar analysis can be performed for the low-strain sample.

The third-order response of the excitonic system can be described by double-sided Feynman diagrams that traverse the available levels for the quantum system.[26] Figure 3(a) shows the level diagram arising from a Hilbert space transform of the electronic states in bulk GaAs including the spin degeneracy and strain splitting. Optical transitions are shown in the circular basis, where $\sigma^-$ and $\sigma^+$ correspond to left and right circular polarization. The resulting transitions are a ground state $|g\rangle$ indicating no optical excitation, HH and LH excitons denoted $|h\rangle$ and $|l\rangle$, and four biexcitons denoted $|hh\rangle$, $|ll\rangle$, $|hl\rangle$ and $|lh\rangle$. In the diagrams, these bound biexcitons are drawn as solid lines below the dashed lines to illustrate the sub-meV bulk biexciton binding energy.[43] Note that the spin quantum number has been omitted for ease of discussion of linearly polarized excitation.

Quantum pathways traversing the level diagram obey time-ordering and phase-matching condition corresponding to rephrasing spectra.[26] Signals must occur from excitation by all three pulses; hence, the language of polarization and population can be employed as a result of successive interactions with odd and even numbers of laser pulses. The first pulse drives a polarization between the ground and an excited state, the second pulse leads to a population (or an excited-state coherence) and the third pulse drives a polarization that radiates as the FWM signal. Figure 3(b) shows ground-state bleaching (GSB) pathways, where the second pulse creates a population in the ground state. Figure 3(c) shows excited-state emission (ESE) pathways, where the second pulse drives a population/polarization in the first excited states (excitonic) manifold. In 2DFTS spectra both GSB and ESE pathways result in diagonal intra-action features for both exciton species and two off-diagonal interaction features that are a result of coupling between the two exciton species. Figure 3(d) shows excited-state absorption (ESA) pathways, where the signal arises from a final population of the excited state and includes a polarization between the excitonic and second-excited (biexcitonic) manifolds. In 2DFTS spectra, ESA contributions can be used to identify the bound biexcitons, because the biexciton binding energy results in a shift in the emission energy relative to the absorption energy. To date 2DFTS has revealed biexcitons features adjacent to the heavy-hole exciton in GaAs quantum wells,[23,24] interfacial quantum dot distributions[44] and upper branch exciton-polaritons in semiconductor microcavities.[45]

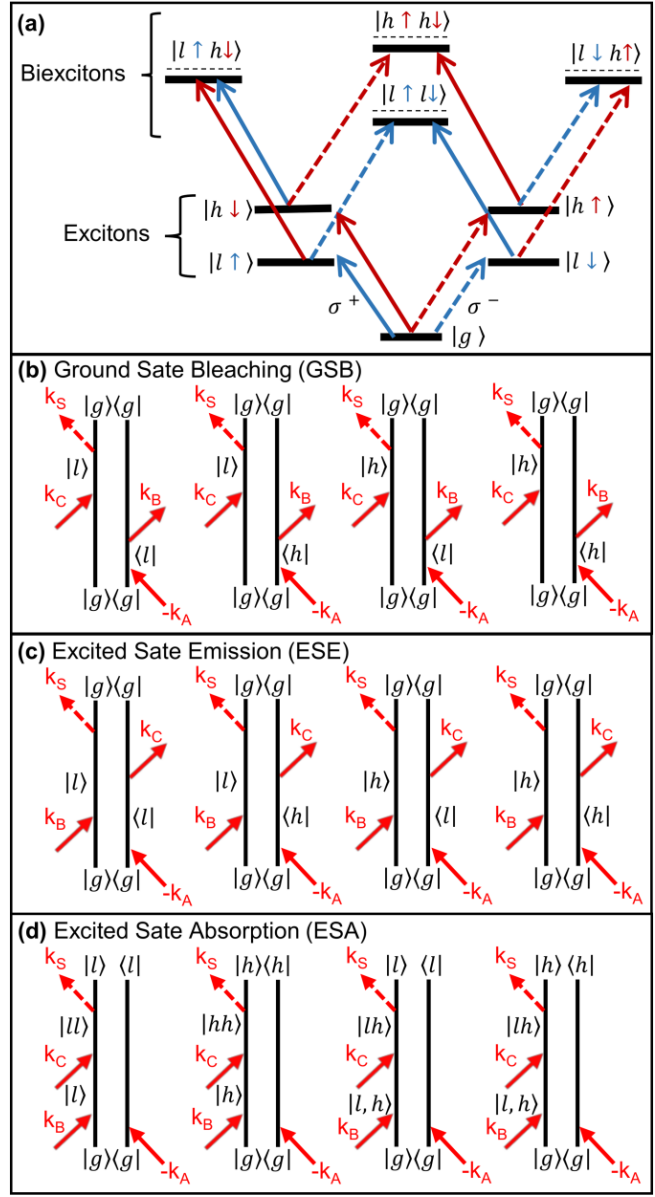

**Fig. 3** (color online) (a) Exciton level diagram for strain-split bulk excitons and biexcitons in GaAs. Double-sided Feynman diagrams illustrate the quantum pathways of (b) ground state bleaching, (c) excited state emission and (d) and excited state absorption.

Figure 4 shows the 2DFTS rephasing spectra for low- (0.08%) and high-strain (0.18%) samples: (a) amplitude and (e) real parts of the XXXX polarized spectra are shown for the low-strain sample; (b) and (f) show the same for XYYX polarization. Similarly, high-strain spectra are show in panels (c), (d), (g) and (h). All spectra are normalized to their strongest feature. First, it is worth discussing the spectral features that are common to both samples. Each spectrum is dominated by four star-like features as described above, namely two diagonal ($X_L$ and $X_H$) and two off-diagonal ($X_L$-$X_H$ and $X_H$-$X_L$) coherences.

In addition, XXXX polarized spectra for both samples show two vertical streaks (denoted $C_L$, and $C_H$), which are coherences resulting from MBI between the electron-hole continuum and the $X_L$ and $X_H$ resonances.[33,46] Selecting XYYX polarization is known to suppress MBI,[47,48] such that in 2DFTS spectra the signals are reduced by a factor of approximately five, the $C_L$ and $C_H$ features are strongly reduced, the off-diagonal features become similar or weaker than the diagonal features,[23,24] and biexciton features become more prominent.[24,44] The latter effect is more pronounced for the high strain sample. For both samples, dispersive line shapes are observed in the real part of the XXXX polarized spectra as a result of the MBI.[22,23] For XYYX polarization, with suppressed MBI, the spectral features become peaked about their center position. For the case of the diagonal features, this means they are centered on the diagonal with positive sign, while off-diagonal features are centered at positions corresponding to absorption into one exciton species and emission from the other with negative sign.

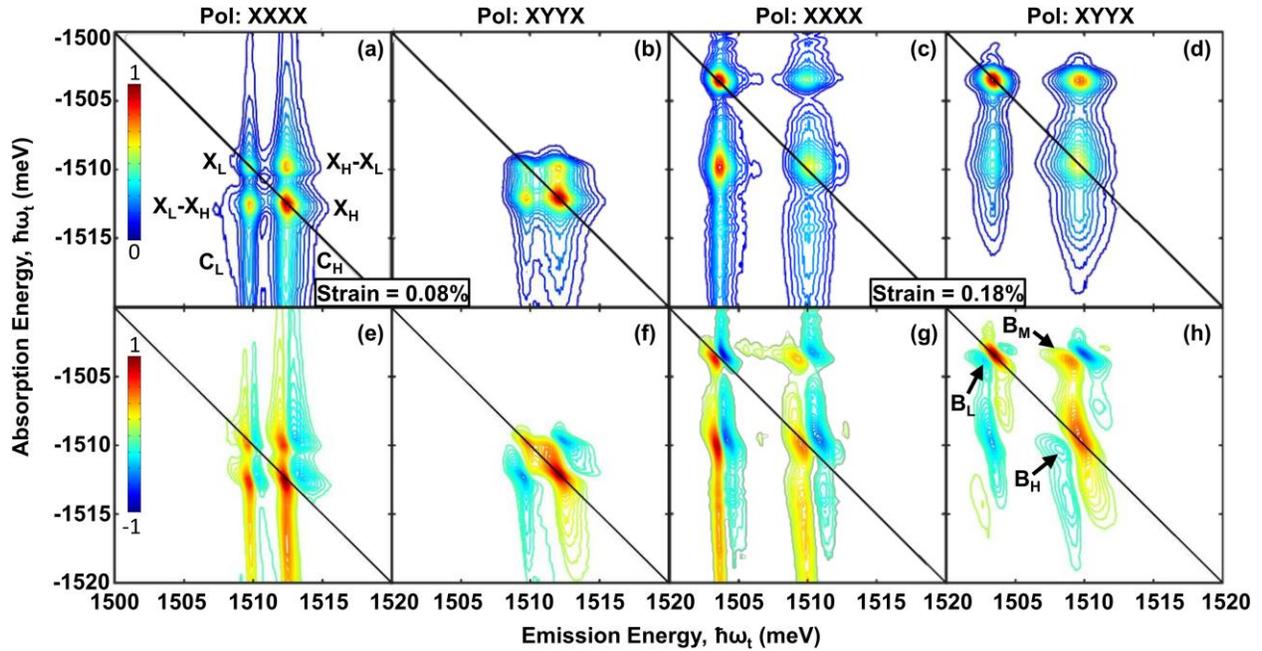

**Fig. 4** (color online) Rephasing 2DFTS spectra for XXXX and XYYX polarization of the low-strain (0.08%) and high-strain (0.18%) GaAs samples. Panels (a)-(d) show the normalized amplitude and panels (e)-(h) show the corresponding real part of the spectra. Diagonal (intra-action) coherences $X_L$ and $X_H$ are the LH and HH excitons respectively, off-diagonal (interaction) coherences $X_L$-$X_H$ and $X_H$-$X_L$ reveal coupling between exciton species. Features $C_L$ and $C_H$ are electron-hole continuum coherences with their respective excitons. Features $B_L$, $B_H$, and $B_M$ correspond to LH, HH, and mixed biexciton coherences in the high-strain sample.

Increasing the strain from 0.08% to 0.18% has several effects on the coherent properties observed in the 2DFTS rephasing spectra. As seen in the linear spectra, the splitting between the two exciton resonances increases with strain. The spectral coordinates {$E_{abs}$, $E_{emi}$} of the four main features $X_L$, $X_H$, $X_L$-$X_H$ and $X_H$-$X_L$ are listed in Table 1 for both low- and high-strain samples. The increased splitting has the effect of making the exciton species more energetically distinct as is evidenced by an alteration in the spectral weights of the four main features of Figs. 4(a) and (c). It can be seen that in the low-strain sample the $X_H$ has larger oscillator strength than $X_L$, and that $X_L$-$X_H$ and $X_H$-$X_L$ are equal in magnitude, with strengths that are approximately the average of $X_L$ and $X_H$. This result is reminiscent of the coherent response of three-level atomic vapors,[49] indicating a symmetric coupling between the two exciton species.[26] On the other hand, in the high-strain sample the lower energy exciton $X_L$ is stronger than $X_H$ and the relative strengths of the off-diagonal features become asymmetric, with $X_L$-$X_H$ increasing by approximately the same amount that $X_H$-$X_L$ decreases. This asymmetry is a signature of the excitation-induced dephasing (EID) and excitation-induced shift (EIS).[22,46] This is reminiscent of the coherent response in lower dimensional GaAs systems and not incoherent energy relaxation, since $T = 100$ fs. The comparison of the two results suggest a greater impact of MBI in the high-strain sample. More study is required to elucidate the role of MBI in the two samples.[33]

Another prominent effect of the increased strain is the presence of new features in XYYX rephasing spectra, which tend to be more sensitive to bound biexciton contributions that arise from ESA pathways; see Fig. 3(d). The real-part of XYYX spectrum in Fig. 4(h) exhibits a strong negative $B_H$ feature. Spectral coordinates {$E_{abs}$, $E_{emi}$} are listed in Table 1, revealing the shift in the emission energy that corresponds to the biexciton binding energy. This contribution arises from the ESA pathway shown second to left in Fig. 3(d) and has been observed in other excitonic systems as discussed above. In addition, new features are observed. The feature at $B_L$ is associated with a LH biexciton and the left most double-sided Feynman diagram in Fig. 3(d). The feature at $B_M$ is associated with a mixed biexciton and the right most double-sided Feynman diagram of the same series. Energetically, the mixed biexciton pathway involves pulse $A^*$ creating a polarization between $|g\rangle$ and $|h\rangle$, pulse $B$ creating a polarization in either $|h\rangle$ or $|l\rangle$, pulse $C$ creating a polarization with $|lh\rangle$ or $|hl\rangle$ and emission releasing a photon with energy $E_H$-$\Delta E_M$ to return the system to a population in $|h\rangle$. From the lateral shifts of the three biexcitonic features determined from the spectral coordinates listed in Table 1, it is estimated that the binding energies are ~1 meV and ~2 meV for the LH and HH biexcitons, which compare well to literature values.[41] Averaging the two pure biexciton binding energies gives a value close to observed mixed biexciton binding energy, $\Delta E_M = 1.2 \pm 0.1$ meV. The presence of the mixed biexciton enhances the $X_H$-$X_L$ feature in the normalized amplitude spectra for this polarization [see Fig. 4(d)], as compared to the EID and EIS enhancement of the $X_L$-$X_H$ feature for XXXX polarization [see Fig. 4(c)]. To the best of the authors' knowledge, mixed biexcitons have been theoretically explored in one-quantum rephasing[26] and two-quantum[27] 2DFTS and experimentally observed in two-quantum spectra,[25,50] but have not been directly observed in one-quantum rephasing spectra. Biexciton contributions are not observed in the low-strain sample, likely because of the relatively small HH/LH splitting which is comparable to the biexciton binding energy.

**Table 1** Spectral coordinates for the exciton and biexciton features in units of meV and with error of ~0.1 meV. Exciton and biexciton features are extracted from XXXX and XYYX spectra respectively.

| Spectral Feature | Low Strain (0.08%) | | High Strain (0.18%) | |
|---|---|---|---|---|
| | $E_{abs}$ | $E_{emi}$ | $E_{abs}$ | $E_{emi}$ |
| $X_L$ | -1509.7 | 1509.7 | -1503.5 | 1503.5 |
| $X_H$ | -1512.4 | 1512.4 | -1510 | 1510 |
| $X_L$-$X_H$ | -1512.4 | 1509.7 | 1510 | 1503.5 |
| $X_H$-$X_L$ | -1509.7 | 1512.4 | -1503.5 | 1510 |
| $B_H$ | - | - | -1510.2 | 1508.2 |
| $B_L$ | - | - | -1503.8 | 1502.8 |
| $B_M$ | - | - | -1503.8 | 1509 |

Star-like line shapes indicate that the exciton coherences have nominal inhomogeneous broadening,[18] as might be expected for bulk excitons where the excitonic Bohr radius is more than an order of magnitude smaller than the GaAs layer thickness. This contrasts with 2DFTS spectra for excitons in low-dimensional semiconductors, where local variations in quantum confinement within the excitation spot lead to significant inhomogeneous broadening and spectral features that are elongated along the diagonal direction. Nonetheless, the degree of inhomogeneous broadening is acquired using the analysis method of Siemens et al.[18] which simultaneously fits the diagonal and cross-diagonal slices of each spectral feature, relating the line shapes to the homogeneous ($\Gamma$) and inhomogeneous ($\sigma$) linewidths.

Table 2 shows the extracted linewidths from the $X_L$ and $X_H$ diagonal features for the four amplitude spectra shown in Fig. 4(a) to (d). The error associated with each value from averaging is approximately 10% or 0.07 meV. Nonetheless, comparisons can be made for the two samples for the same polarization. All spectra show $\Gamma \geq \sigma$ for both excitons, within the experimental uncertainty, indicating that the features are indeed all primarily homogeneously broadened. Moreover, the homogeneous linewidth of $X_L$ is smaller than that for $X_H$ when excited by XXXX polarization, where excitonic linewidths are less complicated. This result is consistent with visual inspection of both linear and 2DFTS measurements. For both polarizations, the homogeneous linewidths of $X_L$ and $X_H$ are more comparable in the low-strain sample than the high-strain sample, consistent with the change in coupling discussed above. Namely, closer spaced energy levels lead to comparable line shapes, as well as comparable amplitudes of the off-diagonal elements. Similar changes in the distinguishability of 2DFTS features has been observed in wedged semiconductor microcavities,[45] where the energy separation between lower and upper branches of the exciton-polaritons can be tuned to vary the interactions strength. For XYYX, biexcitons are expected to be similar strength to excitons, which complicates direct comparison of linewidths extracted from the two polarizations. This may also increase $\Gamma$ for the low-strain sample.

**Table 2** Homogeneous ($\Gamma$) and inhomogeneous ($\sigma$) linewidths, with units of meV and an error of ~10%.

| Strained excitons | | XXXX | | XYYX | |
|---|---|---|---|---|---|
| | | $\Gamma$ | $\sigma$ | $\Gamma$ | $\sigma$ |
| 0.08% | $X_L$ | 0.44 | 0.06 | 0.74 | 0.21 |
| | $X_H$ | 0.46 | 0.27 | 0.59 | 0.33 |
| 0.18% | $X_L$ | 0.29 | 0.34 | 0.29 | 0.27 |
| | $X_H$ | 0.70 | 0.41 | 0.87 | 0.31 |

Turning attention to the inhomogeneous broadening, it can be seen that in all cases $\sigma$ is larger for $X_H$ than for $X_L$, as has been observed in quantum wells.[19] The inhomogeneous broadening is also larger in the high-strain sample, which indicates a larger degree of variation in the microscopic electronic environment within the excitation volume for that sample. Since no variation is expected in the plane of the sample, the inhomogeneous broadening is attributed to strain relaxation in the growth direction ($z$). Strain relaxation typically occurs on length scales larger than the exciton Bohr radius,[51] such that at any one point in the excitation volume the exciton's energy levels are related to the local potential. The lower the biaxial strain the smaller the splitting, the higher the strain the larger the splitting. Hence, a distribution of strain values experienced by the excitons locally in the sample is proportional to the average of the exhibited inhomogeneous broadening contribution to the linewidth in the diagonal direction of a 2DFTS feature. In general, if a number of spectral features are observed this is given as $\sum_N W_{G,N}/N$, where $N$ is the peak index and $W_{G,N}$ is the Gaussian full-width half maximum (FWHM) of the peak, associated with inhomogeneous contribution to the diagonal Voigt profile[18] for the $N^{th}$ spectral feature. For moderate amounts of inhomogeneous broadening, the diagonal FWHM of the $N^{th}$ feature is described as $W_N = 0.5346 W_{L,N} + \sqrt{0.2166 W_{L,N}^2 + W_{G,N}^2}$, where $W_{L,N} = 2\gamma_N$ and $W_G = 2\sqrt{2 \ln 2}\sigma_N$.[52] For $X_L$ and $X_H$ features, the average FWHM is $(W_{G,L} + W_{G,H})/2$, which yields $0.39 \pm 0.1$ meV and $0.88 \pm 0.1$ meV for the low- and high-strain samples respectively, based on $\sigma$ values for XXXX polarization. Employing the same conversion used for the average splitting of the exciton line centers, the distribution in effective strain through the depth of the sample is $\{\Delta e\}_z = (W_{G,L} + W_{G,H})/(4|b|)$ This yields 0.011 % and 0.025 % for the low- and high-strain samples respectively. Finally, if it is assumed that the strain relaxes uniformly over the thickness of each sample, then the effective strain relaxation length for the two samples is ~139 %/cm for the 800 nm thick layer and 841 %/cm for the 300 nm thick layer. These numbers reflect the situation that the thinner sample has larger strain, larger splitting between the excitons, a steeper strain relaxation gradient and larger contribution to inhomogeneous linewidth.

## IV. CONCLUSIONS

Two samples of bulk GaAs attached to sapphire disks experience differing amounts of biaxial tensile strain due to their relaxation over differing volumes upon cooling. Strain splits and shifts the observed HH and LH excitons in optical absorption and 2DFTS, which is used to measure the coherent response. Increasing the strain has several effects on the exciton resonances: (i) the separation between the excitons increases; (ii) the strength of off-diagonal coherences become more asymmetric; (iii) the difference between the HH and LH homogenous linewidths increases as a result of greater energy separation; and (iv) the inhomogeneous broadening increases due to an increase in the strain relaxation gradient in the growth direction for thinner samples.

By focusing on bulk layers that are significantly thick compared to the excitonic Bohr radius, the influence of strain on the exciton and biexciton resonances is probed in the absence of quantum confinement. The application of 2DFTS extracts both the average effective strain and distribution in effective strain experienced by the excitons throughout the excitation volume. Using collinear polarization, the one-quantum excitonic features are enhanced by many-body interactions and dominate the spectra. The spectral weight of the excitons become sensitive to the degree of separation induced by the strain. For small HH/LH separation, the spectral weight of the 2DFTS features is more akin to an atomic vapor than the usual semiconductor nanostructure response. Alternatively, increasing strain allows the spectra to begin exhibiting the effects associated with increased excitation-induced MBI.

It is also shown that biexcitons are more easily observed in structures with large average strain due to the large HH/LH splitting relative to the biexciton binding energy. For cross-linear polarization, no population grating is created and, as a result, signal contributions associated with MBI are suppressed, allowing biexciton contributions to be revealed. Biexcitons are a manifestation of few-body interactions and are also expected to be more prevalent in samples where the HH and LH states are distinct. In addition, the first evidence of a mixed biexciton observed by one-quantum 2DFTS is presented.

Overall, these results show the sensitivity of 2DFTS to strain through interrogation of the coherent response and effects on the spectral line shape and spectral weight of all visible features. In particular, the analysis of the inhomogeneous broadening of bulk excitons reveals strain gradients that are stronger in thinner GaAs layers. This significant change in the coherent response invites studies where the strain can be controlled while measuring EID, EIS and biexciton formation.[33] Information learned here is part of the growing series of studies aimed at using sophisticated optical spectroscopy to unravel the electronic environment inside micro and nanostructures.

**ACKNOWLEDGEMENTS**
We thank Richard P. Mirin, Margaret Dobrowolska, Xinyu Liu, and Jacek K. Furdyna for growing of the GaAs samples. Work at West Virginia University was funded by the West Virginia Higher Education Policy Commission (HEPC.dsr.12.29) and the National Science Foundation (CBET-1233795). Work at Dalhousie University was funded by Natural Sciences and Engineering Research Council of Canada. JMA was supported by the National Science Foundation through the Research Experience for Undergraduates program (DMR-1262075).

\* E-mail: alan.bristow@mail.wvu.edu
† Current address: Department of Physics, Radford University, Radford VA 24142, U.S.A.